\documentclass[letter,traditabstract]{aa}

\usepackage{graphicx}
\usepackage{txfonts}
\usepackage{epsfig}
\usepackage{natbib}
\bibpunct{(}{)}{;}{a}{}{,}

\begin{document}

\title{Detection of the host galaxy of S5 0716+714}

\author{K. Nilsson\inst{1}
\and T. Pursimo\inst{2}
\and A. Sillanp\"a\"a\inst{1}
\and L. O. Takalo\inst{1}
\and E. Lindfors\inst{1,3}
}

\institute{Tuorla Observatory, University of Turku, V\"ais\"al\"antie 20,
FI-21500 Piikki\"o, Finland
\and Nordic Optical Telescope, Apartado 474, E-38700 Santa Cruz de La Palma
Santa Cruz de Tenerife, Spain
\and 
Mets\"ahovi Radio Observatory, TKK, Helsinki University of Technology,
Mets\"ahovintie 114, FIN-02540 Kylm\"al\"a, Finland
}

\date{Received / Accepted}

\abstract{

We have acquired a deep i-band image of the BL Lacertae object
\object{S5 0716+714} while the target was in an low optical state. Due
to the faintness of the nucleus, we were able to detect the underlying
host galaxy. The host galaxy is measured to have an I-band magnitude
of 17.5 $\pm$ 0.5 and an effective radius of (2.7 $\pm$ 0.8)
arcsec. Using the host galaxy as a ``standard candle'', we derive z =
$0.31 \pm 0.08$ (1$\sigma$ error) for the host galaxy of \object{S5
  0716+714}.  This redshift is consistent with the redshift z = 0.26
determined by spectroscopy for 3 galaxies close to \object{S5
  0716+714}.  The effective radius at z = 0.31 would be 12 $\pm$ 4
kpc, which is consistent with values obtained for BL Lac host
galaxies.  An optical spectrum acquired during the same epoch shows no
identifiable spectral lines.

}


\titlerunning{The host galaxy of S5 0716+714}

\maketitle

\section{Introduction}

BL Lacertae objects (BL Lacs) are active galactic nuclei (AGN)
characterized by a featureless nonthermal continuum, high optical and
radio polarization, and variability across the entire electromagnetic
spectrum.  These properties arise from a relativistic jet pointed
almost towards the observer \citep{1979ApJ...232...34B}. The strong
continuum emission from the jet in the optical sometimes presents a
problem when attempting to properly characterize the underlying host
galaxy or measure its redshift, especially since by definition the
emission lines are very weak in BL Lacs. Hence, it is unsurprising
that even after very deep searches \citep[e.g.][]{2006AJ....132....1S}
many BL Lacs remain without a spectroscopically determined redshift.

The redshift is a fundamental property of any extragalactic object,
without which e.g. the luminosity or intrinsic variability properties
of the target cannot be determined.  Furthermore, many BL Lacs have
been recently detected at TeV gamma-rays using ground-based
Cherenkov telescopes MAGIC, HESS, and VERITAS
\citep[e.g.][]{2007arXiv0712.3352H}.  Since VHE gamma-rays can be
absorbed by the interaction with low energy photons of the EBL via
pair production, this opens the possibility of measuring the amount of
extragalactic background light \citep[EBL,
][]{nikishov1962,1992ApJ...390L..49S} in the optical through to the
far infrared, which provides important information about the galaxy
and star formation history.  The absorption depends strongly on the
distance of the source and the energy of the gamma-rays. If the
redshift of the source is known, the VHE gamma-ray spectrum can be
used to derive limits on the EBL
\citep[e.g.][]{2006Natur.440.1018A,2007A&A...471..439M}.  On the other
hand, if the redshift is unknown, the VHE gamma-ray spectra can be
used to set an upper limit to the redshift of the source
\citep[e.g.][]{2007ApJ...655L..13M}.

After its discovery as a bright (S$_{\rm 5 Ghz}$ $>$ 1 Jy)
flat-spectrum ($\alpha \leq -0.5$, S$_{\nu} \propto \nu^{\alpha}$)
radio source, the BL Lac object \object{S5 0716+714} has been studied
intensively at all frequency bands. The object is highly variable with
rapid variations observed from radio to X-ray bands
\citep{1996AJ....111.2187W,2006A&A...451..797O}. The nucleus of
\object{S5 0716+714} is typically bright in the optical
\citep[see e.g.][]{2005AJ....130.1466N}, thus previous attempts to
either characterize its host galaxy \citep{1993A&AS...98..393S,
2000ApJ...532..816U, 2002A&A...381..810P} or to determine its redshift
spectroscopically \citep{1993A&AS...98..393S, 1996MNRAS.281..425M}
have not been successful. Thus, in spite of numerous variability
studies of \object{S5 0716+714}, it has never been possible to
determine reliably e.g. the linear dimensions and luminosities of the
varying components.

\object{S5 0716+714} has also been detected in TeV
gamma-rays \citep{atel1500}. Given that previous estimates
\citep[e.g.][]{1996AJ....111.2187W} place \object{S5 0716+714} at z
$>$ 0.3, this would make the target one of the most distant TeV
sources detected so far. Given the obvious importance of \object{S5
  0716+714} to EBL studies, it is be important to secure an accurate
spectroscopic redshift for \object{S5 0716+714} or at least constrain
the redshift by some other means.

\object{S5 0716+714} is regularly monitored at Tuorla Observatory as
part of the blazar monitoring
program\footnote{http://users.utu.fi/$\sim$kani/1m/index.html}. On
December 17, 2007 we observed \object{S5 0716+714} to go into a fairly
deep optical minimum (R $\sim$ 14.8) and initiated prompt i-band
imaging at the Nordic Optical Telescope (NOT) to detect its host
galaxy. The imaging was performed five days after the minimum and the
results are reported in this Letter.

Throughout this Letter, we use the cosmology $H_0 = 70$ km s$^{-1}$
Mpc$^{-1}$, $\Omega_{M}$ = 0.3 and $\Omega_{\Lambda}$ = 0.7.

\section{Observations and data analysis}

\object{S5 0716+714} was observed at the Nordic Optical Telescope
(NOT), La Palma on Dec 22, 2007. We used the ALFOSC instrument in
imaging mode with a 2k E2V CCD chip of a gain 0.726 e$^-$ ADU$^{-1}$
and readout noise 3.2 electrons.  Twenty-eight exposures of 45 s were
acquired using an i-band interference filter of almost uniform
transmission between 725 and 875 nm, which provided a total exposure
time of 1260 s.  The i-band filter was chosen because it detects light
above the 4000 \AA\ break of the host galaxy even at z = 0.5, which
maximizes the amount of light from the host galaxy.

\begin{figure}
\centering
\includegraphics[width=8.5cm]{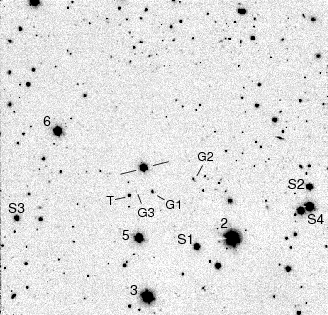}
\caption{\label{kentta} The summed i-band image of \object{S5
0716+714} obtained at the NOT in Dec 22, 2007.  The field size is 3.85
$\times$ 3.7 arcmin, north is up and east is to the left. Object T is
a transient or a very red object not visible in previous images of
\object{S5 0716+714}, most of which were obtained in the R-band.
The other labeled objects are discussed in the text.  }
\end{figure}

The individual images were bias-subtracted and flat-fielded with
twilight flats in the usual way using IRAF\footnote{ IRAF is
  distributed by the National Optical Astronomy Observatories, which
  are operated by the Association of Universities for Research in
  Astronomy, Inc., under cooperative agreement with the National
  Science Foundation.}. A fringe pattern with an amplitude of $\sim$
2.5\% of the sky background was visible in the reduced images. A
fringe correction image was produced by median combining all 28 images
and applying simultaneously a sigma clipping cut to the pixel
intensity values. Since the position of \object{S5 0716+714} on the
CCD was altered between the exposures, this procedure produced a clean
correction image consisting purely of the fringe pattern. After
correction for the fringe pattern, the individual frames were
registered and summed.  The summed image shown in Figure \ref{kentta}
has a FWHM of 0.80 arcsec and the sky background is flat within 0.4\%
over the entire field of view.

The field was calibrated using stars 3, 5, and 6 for which
\cite{1997A&A...327...61G} published Cousins I-band magnitudes.
Unfortunately, all three stars were saturated in their cores ($r$ $<$
0\farcs4 - 0\farcs8), and could not be used directly for calibration
purposes. Instead, we used a 45 s i-band exposure of the field
acquired immediately before the \object{S5 0716+714} sequence. The
FWHM in this image is significantly worse than in the the \object{S5
  0716+714} sequence and stars 3, 5, and 6 are not saturated. Using
these stars we first determined the zero point of the 45 s exposure.
All three stars  gave consistent results, the maximum difference
between the zero points was 0.047 mag, consistent with the errors of
the I$_{\rm C}$ magnitudes 0.04-0.05 mag in
\cite{1997A&A...327...61G}.  Using this zero point, we determined the
I$_{\rm C}$ magnitudes of stars S1, S2, and S3 and from these three
stars the zero point of the combined long exposure. The uncertainly in
the zero point was dominated by the uncertainty in the I$_{\rm C}$
magnitudes of stars 3, 5, and 6 and by the color effects between our
i-band filter and the I$_{\rm C}$ filter. We estimate the former
contribution to the uncertainty to be 0.03 mag.

To estimate the magnitude of the color effects, we used the
established relation between I$_{\rm C}$ and and SDSS i given by
Lupton (2005) in the SDSS web pages
\footnote{http://www.sdss.org/dr4/algorithms/sdssUBVRITransform.
html\#Lupton2005}: $I = i - 0.3780(i-z) -0.3974$.  The SDSS i has
approximately the same width as our i-band filter, but the central
wavelength is $\sim$ 50 nm lower. For stars 3, 5, and 6, the $i-z$ color
is $\sim$ -0.01 -- 0.06. Assuming stars S1-S3 have similar colors and
since an elliptical galaxy at z = 0.3-0.5 has a color $i-z \sim
0.4-0.5$, the color effects between the calibration stars and the host
galaxy of \object{S5 0716+714} may cause an error of up to 0.2 mag in
SDSS i. Since our filter is close to SDSS i, we expect the color
effects to be similar in value. Since the color errors dominate the
total error of the zero point, we assign a formal error of 0.2 mag to
the magnitude zero point.

The host galaxy was analyzed by fitting a two-dimensional surface
brightness model to the light distribution of \object{S5 0716+714}.
Details of this process can be found in \cite{1999PASP..111.1223N}.
In short, the model consists of two components, an unresolved core
representing the BL Lac nucleus and a de Vaucouleurs profile
representing the host galaxy. Previous imaging
\citep[e.g.][]{2000ApJ...532..816U} has shown that the de Vaucouleurs
profile describes well the surface brightness profiles of BL Lac host
galaxies, although deviations from this law have also been observed
\citep[e.g.][]{2003A&A...400...95N}. For our purposes, it is
sufficient to know that BL Lac host galaxies are bulge-dominated
systems and no BL Lacertae object has ever been reliably associated
with a disk type host.

The model is described by 5 parameters: position ($x$,$y$), core
magnitude $m_C$, host galaxy magnitude $m_H$, and host galaxy
effective radius $r_{\rm eff}$. The model parameters are adjusted
using an iterative Levenberg-Marquardt loop, until the minimum value
of the $\chi^2$ statistic between the model and the data is
found. Only pixels within 9\farcs5 from the center of \object{S5
  0716+714} were included in the fit and pixels influenced by an I =
20.8 mag nearby companion 4\farcs2 west of \object{S5 0716+714} were
excluded.

The model is convolved by the observed PSF, determined from stars in
the vicinity of \object{S5 0716+714}. Since stars 5 and 6 were
saturated in their cores (r $<$ 0.4 arcsec; the peak counts of
\object{S5 0716+714} remained under 58~000 ADUs in all images.), we
constructed the PSF from stars 5 and S1 by joining smoothly the outer
parts of star 5 with the inner part of S1 (hereafter we refer to this
PSF as PSF1).  A second PSF was constructed in a similar way from
stars 6 and S2 (PSF2) to study the effect of PSF variability over the
field of view.

Since the host galaxy is faint compared to the nucleus, the results
are highly dependent on the accuracy of the PSF and its stability over
the field of view. We analyzed the PSF stability by extracting the
surface brightness profiles of stars 2, 3, 5, 6, S1, and S2 and
scaling them to the same magnitude. In the lower panel of
Fig. \ref{prof}, we show the surface brightness profiles of these
stars relative to star 5, which is equal to PSF1 at $r >$ 0\farcs4.
We note that in this representation star 5 appears as a horizontal
line.  The surface brightness profiles of the stars agree to within
0.2 mag all the way to the outer fitting radius of 9\farcs5. The rms
scatter between the surface brightness profiles is 0.8\% close to the
center and increases to $\sim$ 10\% at the outer fitting radius. Part
of this increase is undoubtedly due to the increase in noise in
measuring the surface brightness at the faint outer wings of the
stars.  However, we assume conservatively that the increase is
entirely due to PSF variability and represent it by a parabolic
expression
$$
\sigma_{\rm PSF}(r) = 0.008 + 4 \times 10^{-5} \cdot r^2\ ,
$$
where $\sigma_{\rm PSF}(r)$ is the uncertainty in the PSF relative to
the intensity at radius $r$ (pixels) from the PSF center. We then
compute the expected variance of a pixel $\sigma^2$ with intensity
$I$ (ADU) and distance $r$ from the center of \object{S5 0716+714} from
the expression
$$
\sigma^2 = \frac{G*I+R^2}{G^2} + [\sigma_{\rm PSF}(r) * I]^2\ ,
$$
where $G$ is the effective gain and $R$ is the effective readout noise,
and compute the $\chi^2$ from
$$
\chi^2 = \sum_i \frac{(I_i - M_i)^2}{\sigma_i^2}\ ,
$$
where $M$ is the model intensity and the summation is over all
unmasked pixels within the fitting radius.

In addition to the i-band images, two spectra with an exposure time of
900 s were obtained using grism \#4 of ALFOSC, which covers the
wavelength range 3200-9100 \AA. The spectral resolution achieved using
the 1\farcs3 slit was 21\AA. Due to the effect of the secondary
spectrum and fringing longwards of 6000 \AA, we studied only the
wavelength range 4000-6000 \AA. The summed spectrum has a
signal-to-noise ratio (S/N) of $\sim$ 250, but the spectrum remains
featureless.

\section{Results}

\begin{figure}
\includegraphics[width=9cm]{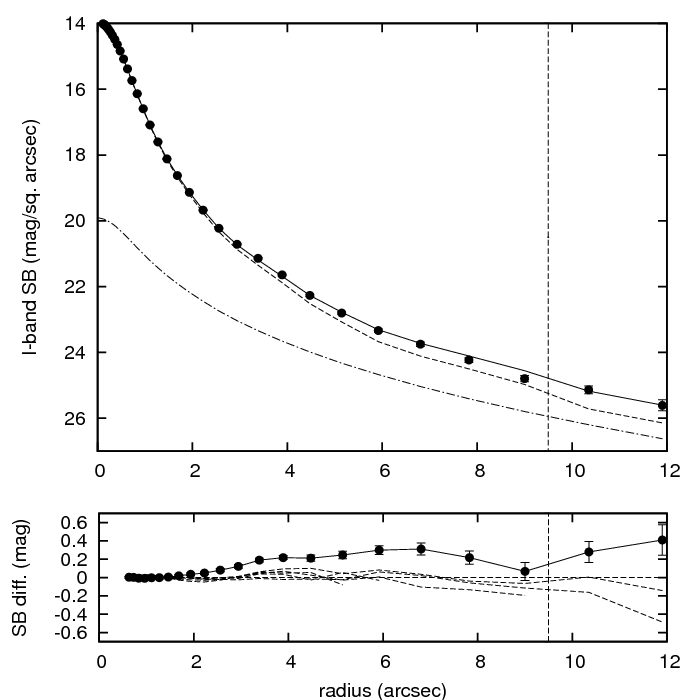}
\caption{\label{prof} Upper panel:
The I-band surface brightness profile of
\object{S5 0716+714} (filled symbols). The solid line shows the
core + host galaxy model, the dashed line the core component and the
dot-dashed line the host galaxy component. The vertical dashed line shows
the outer radius of the model-fitting region. Lower panel:
The surface brightness profiles of \object{S5 0716+714} (symbols and solid
line) and stars 2, 3, 5, 6, S1, and S2 (dashed lines) relative to star 5.
Each star is traced out to a radius at which the surface brightness can
be determined to higher accuracy than 10\%.}
\end{figure}

Table \ref{tulokset} indicates the results of the model fits, and the
upper panel of Fig. \ref{prof} shows the one-dimensional surface
brightness profile of \object{S5 0716+714} and the model profiles. The
I-band magnitude of the nucleus translates into R $\sim$ 14.2 using
(R-I) = 0.5 \citep{2006A&A...450...39G,2006MNRAS.366.1337S}, so the
nucleus had brightened by $\sim$ 0.6 mag in the R-band from the deep
minimum on Dec 17th.

The host galaxy is faint (I = 17.5) compared to the nucleus and never
exceeds the core brightness at any radius, but is visible as a small
excess in the surface brightness profile of \object{S5 0716+714} (see
Fig. \ref{prof}).  Since PSF1 was constructed from stars closer to
\object{S5 0716+714} than PSF2, it is unsurprising that PSF1 provides
a more accurate description of the data (smaller $\chi^2$) than PSF2.
The fit with core + galaxy reduces the $\chi^2$ of the fit, but none
of the fits provide formally satisfactory fits, most likely due to
higher-order PSF variability unaccounted for in our radial PSF
variability model.

\begin{table}
\caption{\label{tulokset} Model fit results. PSF1 was derived from stars
5 and S1 in Fig. \ref{kentta} and PSF2 from stars 6 and S2.}
\centering
\begin{tabular}{lccccc}
\hline
\hline
Model & PSF & I$_{\rm core}$ & I$_{\rm host}$ & $r_{\rm eff}$ & $\chi^2$/dof\\
\hline
Core         & PSF1 & 13.74 &       &          & 1.57\\
Core + De V. & PSF1 & 13.77 & 17.47 & 2\farcs7 & 1.31\\ 
Core         & PSF2 & 13.73 &       &          & 1.95\\
Core + De V. & PSF2 & 13.79 & 17.09 & 1\farcs8 & 1.41\\
\hline
\end{tabular}
\end{table}

We completed several tests to ensure that the observed excess is real
and not due to PSF variability over the field of view. Firstly, we
compared the scaled surface brightness profiles of stars 2, 3, 6, S1,
and S2 and \object{S5 0716+714} with the surface brightness profile of
star 5 (see the lower panel of Fig. 2). There is a clear excess around
\object{S5 0716+714} that exceeds the rms scatter of stellar profiles
by a factor of 3-6 at $r >$ 2\arcsec. The excess is clearly above any
PSF variability observed over the field of view.

Secondly, although the fits with PSF2 provide a slightly worse fit
than with PSF1, the host galaxy component is still present with
similar brightness and effective radius as with PSF1. Also this test
indicates that PSF variability does not affect the results
significantly. Finally, we fitted core + host galaxy models to stars
6, S2, and S4, which are similar in brightness to \object{S5
  0716+714}. Using PSF1, the fit converged towards a solution $r_{\rm
  eff} \rightarrow 0$ in all three cases, i.e. no host galaxy
component was detected in targets known to be unresolved.  Based on
the three test mentioned above, we conclude that the excess around
\object{S5 0716+714} is real and we have detected the host galaxy.

Knowing the host galaxy magnitude, we can estimate the redshift of
\object{S5 0716+714}, albeit with rather large margins.
\cite{2005ApJ...635..173S} demonstrated that the distribution of BL
Lac host galaxy absolute magnitude $M_R$ is almost Gaussian with an
average of -22.8 and $\sigma = 0.5$, using the same cosmology as we do
here, and BL Lac host galaxies can therefore be used as a ``standard
candle'' to estimate distances. Before using their method, we first
have to transform the apparent I-band magnitude of the host galaxy to
apparent R-band magnitude. We correct the host galaxy magnitude I =
17.47 for the galactic absorption by assuming $A_I = 0.06$
\citep{1998ApJ...500..525S}.  Since the observed $R -I$ color depends
on redshift, we have to determine the redshift by iteration, starting
from z = 0 and using Eq. (2) in \cite{2005ApJ...635..173S} and typical
elliptical galaxy colors from \cite{1995PASP..107..945F}. This
iteration yields R = 18.3 and z = 0.31 in the adopted cosmology.

To compute the uncertainty in the derived redshift, we use the
difference between the fits with PSF1 and PSF2 as an estimate for the
1$\sigma$ fitting uncertainty of the host galaxy magnitude,
i.e. $\sigma_{I\ host}$ = 0.4 mag. Summing this in quadrature with
the error of the zero point (0.2 mag), the total uncertainty in
the I-band magnitude of the host is 0.45 mag. Performing the redshift
iteration at $I_{\rm host} + 0.45$ mag and $I_{\rm host} - 0.45$ mag
yields an error of $\pm$0.5 mag in the R-band magnitude of the host
and $\pm$0.06 for z. By adding to this in quadrature the inherent
uncertainty in the method \citep[$\Delta z =
  0.05$,][]{2005ApJ...635..173S}, we derive the final error in z to be
0.08.

Finally, we mention that the effective radius of the host galaxy
2\farcs7 $\pm$ 0\farcs9 translates into 12$\pm$4 kpc at z = 0.31,
which is consistent with typical values found for BL Lac host
galaxies.

\section{Discussion}

Our measurement of redshift does not significantly differ from the
value z = 0.26 obtained for three galaxies close to \object{S5
  0716+714} using spectroscopy \citep[][galaxies G1-G3 in Fig.
  \ref{kentta}]{1993A&AS...98..393S,bychkova2006}.  Both the host
galaxy and nearby environment therefore have properties indicative of
an object at z = 0.26.  The nearby companion 4\farcs2 W of \object{S5
  0716+714} remains unresolved in our image.

There have been four  previous attempts to derive the host galaxy
properties of \object{S5 0716+714} with varying success.
\cite{1993A&AS...98..393S} found the object to be unresolved in
a 500 s R-band image obtained under 1.1 arcsec seeing at the Calar
Alto 3.5 m telescope. \cite{2000ApJ...532..740S} obtained a 614 s HST
image through the F702W filter and found the surface brightness
profile to be consistent with the PSF and derived an upper limit
R $>$ 20.0 for the host galaxy. \cite{2002A&A...381..810P} found
that \object{S5 0716+714} was unresolved in a 2120 s R-band exposure
at the NOT, but the PSF was determined partly from a different image
and the PSF shape was uncertain. Finally, \cite{bychkova2006} reported
a detection of excess radiation in the wings of \object{S5 0716+714}
in VRI images of exposure times between 600 and 1500 s obtained with
the 6 m SAO telescope.  They described the excess as very large
(15\arcsec $\times$ 7\arcsec) and elliptical in shape.

\cite{bychkova2006}, hereafter B06, did not perform model fitting to
their image but rather derived the host galaxy properties from
PSF-subtracted images.  Since neither surface brightness values nor
the FWHM were provided by B06, it is difficult to assess if the large
extent of the host galaxy reported in B06 is consistent with our host
galaxy magnitude and effective radius.  However, they indicated that I
$\sim$ 15.1 for the extended emission, far too bright to be consistent
with the faint host galaxy that we observe in our i-band image.

The upper limit $R > 20.0$ derived by \cite{2000ApJ...532..740S},
hereafter S00, may at first glance appear to contradict our estimate
for the host galaxy of R = ($18.3 \pm 0.5$). Based on this upper
limit, \cite{2005ApJ...635..173S} inferred that z $>$ 0.52 for
\object{S5 0716+714}. The brightness of the nucleus of \object{S5
  0716+714} was the same in S00 (R = 14.18) as here (R $\sim$ 14.2),
so the difference cannot be explained by a greater core dominance in
one of our studies. However, the upper limits in
\cite{2000ApJ...532..740S} are based on statistical errors and do not
take into account possible systematic errors. Consequently, there
might be considerable uncertainty in the derived upper limit and the
two results could be consistent within errors.

We finally note that our redshift estimate is based on the assumption
that the host galaxy of \object{S5 0716+714} has an average
luminosity. The failed previous attempts to detect the host galaxy
suggest that the host galaxy may actually be below average luminosity
and the redshift consequently lower that 0.31. We note that the
scatter in the BL Lac host galaxy luminosities is already included in
our redshift error since this scatter is included in the 0.05
uncertainty of the method by \cite{2005ApJ...635..173S}. Thus, within
a margin of error of 2$\sigma$, the redshift could be in the range of
0.15 to 0.47.

\section{Conclusions}

A deep i-band image of the BL Lacertae object \object{S5 0716+714}
acquired at the Nordic Optical Telescope while the target was in an
low optical state has enabled us to detect the underlying host
galaxy. The host galaxy has an I-band magnitude of 17.5 $\pm$ 0.5 and
an effective radius of (2.7 $\pm$ 0.8) arcsec. Using the host galaxy
as a ``standard candle'' and a method proposed by
\cite{2005ApJ...635..173S}, we derive a redshift of z = $0.31 \pm
0.08$ (1$\sigma$ error) for the host galaxy of \object{S5
  0716+714}. This redshift is consistent with the redshift z = 0.26
determined by spectroscopy for 3 galaxies close to \object{S5
  0716+714}.  The corresponding effective radius of the host galaxy
at z = 0.31 would be $12 \pm 4$ kpc, also consistent with the values
obtained for BL Lac host galaxies in other studies. An optical
spectrum obtained at the same epoch shows a featureless continuum with
no identifiable spectral lines.

\begin{acknowledgements}

The authors thank Jochen Heidt for useful discussions during the
preparation of the manuscript. These data are based on observations
made with the Nordic Optical Telescope, operated on the island of La
Palma jointly by Denmark, Finland, Iceland, Norway, and Sweden, in the
Spanish Observatorio del Roque de los Muchachos of the Instituto de
Astrofisica de Canarias.  The data presented here have been taken
using ALFOSC, which is owned by the Instituto de Astrofisica de
Andalucia (IAA) and operated at the Nordic Optical Telescope under
agreement between IAA and the NBIfAFG of the Astronomical Observatory
of Copenhagen.

\end{acknowledgements}

\bibliographystyle{aa}

\bibliography{astroph.bib}

\end{document}